\pdfoutput=1
\documentclass[11pt]{article}
\usepackage{footmisc}
\input{template_core/color_tokens}
\input{template_core/typography}
\input{template_core/layout}
\usepackage{microtype}
\usepackage{amsmath,amssymb,amsfonts,mathtools}
\usepackage{bm}
\usepackage{nicefrac}
\allowdisplaybreaks
\usepackage{graphicx}
\graphicspath{{figures/}{./}}
\usepackage{booktabs}
\usepackage{multirow,makecell,array}
\usepackage{subcaption}
\usepackage{algorithm}
\usepackage{algpseudocode}
\usepackage{enumitem}
\setlist[itemize]{leftmargin=2.2em,itemsep=2pt,topsep=2pt}
\setlist[enumerate]{leftmargin=2.2em,itemsep=2pt,topsep=2pt}
\usepackage{tikz}
\usetikzlibrary{positioning,calc,arrows.meta,shapes.geometric}
\usepackage[round,sort&compress]{natbib}
\usepackage{url}
\usepackage{hyperref}
\hypersetup{colorlinks=true,linkcolor=PaperLink,citecolor=PaperCitation,urlcolor=PaperUrl,breaklinks=true,bookmarksnumbered=true}
\usepackage{bookmark}
\usepackage[capitalise,nameinlink,noabbrev,sort&compress]{cleveref}
\numberwithin{equation}{section}

\providecommand{\PaperShortTitle}{Draft Paper}
\RequirePackage{titlesec}
\RequirePackage{fancyhdr}
\titleformat{\section}{\Large\PaperHeadingFont\bfseries\color{PaperAccentDark}}{\thesection}{0.7em}{}
\titleformat{\subsection}{\large\PaperHeadingFont\bfseries\color{PaperNavigation}}{\thesubsection}{0.7em}{}
\titleformat{\subsubsection}{\normalsize\PaperHeadingFont\bfseries\color{PaperNavigation}}{\thesubsubsection}{0.7em}{}
\titlespacing*{\section}{0pt}{3.0ex plus 0.5ex minus 0.2ex}{1.2ex}
\titlespacing*{\subsection}{0pt}{2.2ex plus 0.4ex minus 0.2ex}{0.8ex}
\titlespacing*{\subsubsection}{0pt}{1.6ex plus 0.3ex minus 0.2ex}{0.6ex}
\titleformat{\paragraph}[runin]{\normalsize\PaperBodyFont\bfseries\color{PaperInk}}{}{0pt}{}
\titlespacing*{\paragraph}{0pt}{1.4ex plus 0.4ex minus 0.2ex}{0.9em}
\RequirePackage[font=small,labelfont={bf,sf,color=PaperAccent},labelsep=colon]{caption}

\newcommand{\PaperHeaderBaselineShift}{1pt}
\renewcommand{\headrulewidth}{0.45pt}
\renewcommand{\headrule}{\hbox to\headwidth{\color{PaperGold}\leaders\hrule height \headrulewidth\hfill}}
\fancypagestyle{plain}{\fancyhf{}\fancyfoot[C]{\small\PaperHeadingFont\color{PaperNavigation}\thepage}\renewcommand{\headrulewidth}{0pt}}

\input{template_core/theorem_environments}
\crefname{theorem}{Theorem}{Theorems}\Crefname{theorem}{Theorem}{Theorems}
\crefname{lemma}{Lemma}{Lemmas}\Crefname{lemma}{Lemma}{Lemmas}
\crefname{proposition}{Proposition}{Propositions}\Crefname{proposition}{Proposition}{Propositions}
\crefname{corollary}{Corollary}{Corollaries}\Crefname{corollary}{Corollary}{Corollaries}
\crefname{conjecture}{Conjecture}{Conjectures}\Crefname{conjecture}{Conjecture}{Conjectures}
\crefname{fact}{Fact}{Facts}\Crefname{fact}{Fact}{Facts}
\crefname{definition}{Definition}{Definitions}\Crefname{definition}{Definition}{Definitions}
\crefname{assumption}{Assumption}{Assumptions}\Crefname{assumption}{Assumption}{Assumptions}
\crefname{example}{Example}{Examples}\Crefname{example}{Example}{Examples}
\crefname{problem}{Problem}{Problems}\Crefname{problem}{Problem}{Problems}
\crefname{openproblem}{Open Problem}{Open Problems}\Crefname{openproblem}{Open Problem}{Open Problems}
\crefname{remark}{Remark}{Remarks}\Crefname{remark}{Remark}{Remarks}
\crefname{claim}{Claim}{Claims}\Crefname{claim}{Claim}{Claims}
\crefname{algorithm}{Algorithm}{Algorithms}\Crefname{algorithm}{Algorithm}{Algorithms}

\DeclarePairedDelimiterX{\inner}[2]{\langle}{\rangle}{#1,#2}

\newcommand{\E}{\mathbb{E}}
\newcommand{\Prob}{\mathbb{P}}
\newcommand{\ind}{\mathbf{1}}

\newcommand{\CompRel}{\mathsf{CompRel}}
\newcommand{\Perm}{\mathsf{S}}
\newcommand{\Var}{\operatorname{Var}}
\newcommand{\Unif}{\operatorname{Unif}}
\newcommand{\Dom}{\operatorname{dom}}
\newcommand{\Img}{\operatorname{im}}
\newcommand{\Zset}{\mathcal{Z}}
\newcommand{\Hset}{\mathcal{H}}
\newcommand{\Ccoll}{C_{\mathrm{coll}}}
\newcommand{\Uall}{\mathcal{U}}
\newcommand{\Trans}{\mathsf{Trans}}
\newcommand{\Qpass}{Q_{\mathrm{pass}}}
\newcommand{\bact}{b_{\mathrm{act}}}
\newcommand{\Zge}{\mathbb{Z}_{\ge 0}}
\newcommand{\defeq}{\coloneqq}

\usepackage{authblk}

\renewcommand{\PaperShortTitle}{Mixed-Profile Permutation Iterates}
\title{Mixed-Profile Recovery of Hidden Permutation Iterates}
\author{Mingda Wan, Deyuan Liu, and Yihao Hu}
\affil{}
\date{}
\hypersetup{
  pdftitle={Mixed-Profile Recovery of Hidden Permutation Iterates},
  pdfauthor={Mingda Wan, Deyuan Liu, and Yihao Hu}
}

\makeatletter
\def\@author{%
{\normalsize\color{PaperInk}\centering
Mingda Wan$^{*}$\par}
\vspace{0.20em}
{\normalsize\color{PaperInk}\centering
\begin{minipage}[t]{0.22\linewidth}\centering
Deyuan Liu$^{*}$
\end{minipage}\hspace{0.06\linewidth}
\begin{minipage}[t]{0.22\linewidth}\centering
Yihao Hu$^{*}$
\end{minipage}\par}
\vspace{0.08em}
{\small\color{PaperMutedInk}Westlake University\hspace{0.55em}\ttfamily\{wanmingda, liudeyuan, huyihao\}@westlake.edu.cn\par}}

\renewcommand{\subsectionmark}[1]{}
\newcommand{\PaperTopSection}{%
  \ifx\@empty\topmark
    \expandafter\@rightmark\topmark\@empty\@empty
  \else
    \rightmark
  \fi
}
\newcommand{\PaperRightHeader}[1]{\raisebox{\PaperHeaderBaselineShift}{\small\PaperHeadingFont\color{PaperMutedInk}\nouppercase{#1}}}
\makeatother

\begin{document}
\makeatletter
\maketitle
\vspace{-0.5em}
\begin{abstract}
For a uniformly random permutation $\pi$ on $[N]$, a learner observes $q$ i.i.d.\ labeled edges and
then one nonadaptive batch of at most $b$ membership queries, and must output $m$ distinct-start
pairs $(R_j,\pi^k(R_j))$. On the mixed profile $m=\lfloor N^\gamma\rfloor$,
$h=\lfloor N^\delta\rfloor$, and $b=m-h$, with fixed $k\ge2$ and $0<\delta<\gamma<1$, we prove a
sharp $0$--$1$ threshold at $Q(N)=\max\{(mN^{k-2})^{1/(k-1)},(hN^{k-1})^{1/k}\}$.
For every fixed $\varepsilon>0$, an explicit learner succeeds with probability $1-o(1)$ above
$(1+\varepsilon)Q(N)$, and every learner succeeds with probability $o(1)$ below
$(1-\varepsilon)Q(N)$. The same sharp statement holds for independent budgets $b=O(m)$ when
$b\ge m$ or $m-b=\Omega(m)$. The upper bound concentrates two passive prefix populations. The lower bound tracks batch
collisions, queries whose images land back in the revealed domain, in a deterministic walk-capacity
inequality, then applies a last-fresh-edge posterior bound. Thus the zero-collision union bound can
fail before the sample threshold changes. A thin independent-budget regime remains open, where
nonterminal collisions can contribute a comparable number of certificates. The results are
information-theoretic and restricted to one nonadaptive batch; they do not address generalization,
computation, total-variation graph sampling, or multi-round adaptivity.

\end{abstract}
\begingroup
\renewcommand{\thefootnote}{\textasteriskcentered}
\footnotetext{Equal contribution.}
\endgroup
\section{Introduction}
\label{sec:introduction}

A membership query to a hidden permutation reveals one directed edge, but certifying the iterate
$\pi^k(r)$ requires a chain of $k$ compatible edges. Passive examples and one nonadaptive batch
interact in both directions: the passive sample identifies useful query locations, and batch answers
can land back in the revealed domain and complete additional chains. We study the passive sample
scale at which a learner can certify a growing list of distinct-start iterates.

Consider depth $3$ with $h=\lfloor N^{1/5}\rfloor$ query-free starts,
$m=\lfloor N^\gamma\rfloor$ outputs for $\gamma\in(1/5,1)$, and batch budget $b=m-h$. Without
queries, every output requires a passive $3$-prefix in the observed domain $D$, whose expected
supply is $(1-o(1))q^3/N^2$. This forces $q\asymp(mN^2)^{1/3}=N^{(2+\gamma)/3}$. With $b$ queries,
the learner can instead complete $b$ passive $2$-prefixes by one query each and cover only $h$
outputs from passive $3$-prefixes. The two populations meet their quotas at
$Q_2(N)=(mN)^{1/2}=N^{(1+\gamma)/2}$ and $Q_3(N)=(hN^2)^{1/3}=N^{11/15}$. Their maximum therefore
has a plateau followed by an exponent kink at $\gamma=7/15$.

A natural converse conditions on a collision-free batch: no new query $u\in U$ has image in
$D\cup U$, so each query completes at most one passive prefix. In this example, the union bound
$b(q+b)/(N-q)$ ceases to vanish at $\gamma=4/15$, strictly before the supply kink at $7/15$. This
leaves two questions. Do collisions alter the sample threshold? And can one identify the exact
leading constant, rather than only its power of $N$?

The first obstruction is combinatorial. A single colliding query can complete several walks, so the
zero-collision event is too restrictive. \Cref{lem:collision-capacity} partitions every revealed
$k$-walk by the first index $J$ at which it visits a new query vertex. Passive walks and
query-terminal walks are bounded by the passive $(k-1)$-walk count, while a nonterminal query forces
a collision and contributes at most $(k-1)\Ccoll$ walks by injectivity. The resulting deterministic
capacity inequality
\[
W_k(M)\le\min\bigl(H_k(D)+b,\,H_{k-1}(D)\bigr)+(k-1)\Ccoll
\]
requires only $\E[\Ccoll]=o(m)$ or $o(h)$ on the two branches.

The second obstruction is posterior uncertainty. Even when fewer than $m$ walks are revealed, the
learner selects its reported starts after seeing the transcript. Fixing the learner's seed first
(\Cref{lem:posterior-completion}) makes the first unforced reported start transcript-measurable.
\Cref{lem:last-fresh} then bounds its conditional endpoint probability by $O_k(1/N)$ through a
sequential last-fresh-edge decomposition.

For the upper bound, preserving the leading constant one requires a near-full passive domain
(\Cref{lem:domain-reduction}) and second-moment concentration of both prefix populations
(\Cref{lem:prefix-moments}). These ingredients yield the sharp threshold in
\Cref{thm:two-branch}, rather than a constant-factor approximation.

\paragraph{Contributions.}
\begin{enumerate}[label=\textup{(\roman*)}]
\item \textbf{Sharp mixed-profile threshold.} \Cref{thm:two-branch} proves that for fixed $k\ge2$
and $0<\delta<\gamma<1$, with $m=\lfloor N^\gamma\rfloor$, $h=\lfloor N^\delta\rfloor$, and
$b=m-h$, success changes from $o(1)$ to $1-o(1)$ across
$Q(N)=\max\{(mN^{k-2})^{1/(k-1)},(hN^{k-1})^{1/k}\}$ at leading constant one.
\Cref{cor:passive-baseline} gives the sharp passive-only scale $\Qpass(N)=(mN^{k-1})^{1/k}$.
\item \textbf{Independent budgets outside a thin boundary.} \Cref{thm:decoupled} proves the same
sharp transition when $m\to\infty$, $m=o(N)$, and $b=O(m)$, provided $b\ge m$ or $m-b=\Omega(m)$.
\item \textbf{Nonterminal collisions and query-only recovery.} \Cref{prop:k2-counterexample}
constructs a $k=2$ learner that succeeds with probability $1-o(1)$ below the naive two-scale value
inside the thin boundary. \Cref{lem:query-only} shows that a uniform batch alone recovers $m$
iterates with high probability when $b=o(N)$ and $b^k/N^{k-1}\gg m$.
\end{enumerate}

\paragraph{Organization.}
\Cref{sec:preliminaries} defines the transcript and walk counts. \Cref{sec:results} states the sharp
threshold and its phase diagram. \Cref{sec:proofs} develops the complete proof chain.
\Cref{sec:decoupled} treats independent budgets and the thin-boundary counterexample.
\Cref{sec:related} separates neighboring access models, and \Cref{sec:discussion} states the open
boundaries. \Cref{app:checks} supplies the algebraic lemmas, the explicit packing procedure, and an
exact finite-instance consistency check.

\section{Model and Notation}
\label{sec:preliminaries}

Fix an integer $k\ge2$. Write $[N]\defeq\{1,\dots,N\}$, let $\Perm_N$ be the symmetric group on
$[N]$, and let $(x)_a=x(x-1)\cdots(x-a+1)$ denote the falling factorial, with $(x)_0=1$. A partial
injection $M$ is identified with its graph $\{(x,M(x)):x\in\Dom(M)\}$, and $M\subseteq\pi$ means
$M(x)=\pi(x)$ on $\Dom(M)$. The hidden permutation $\pi$ is uniform on $\Perm_N$.

\subsection{Passive edges, one batch, and list recovery}
\label{sub:access}

The learner receives $q\in\Zge$ i.i.d.\ labeled edges $(X_i,\pi(X_i))$, where
$X_i\sim\Unif([N])$ independently of $\pi$. Put $X=(X_1,\dots,X_q)$, $D=\{X_1,\dots,X_q\}$, and
$P=\pi|_D$. Conditional on $|D|=d$, the set $D$ is a uniform $d$-subset independent of $\pi$.

The learner's seed $\omega$ takes values in a measurable space and is independent of $(\pi,X)$;
batch-selection and output maps are measurable in $\omega$. After observing $(X,P)$, the learner
commits to a batch $\Uall\subseteq[N]$ with $|\Uall|\le b$ before seeing any answer. Let
$U=\Uall\setminus D$, $M=\pi|_{D\cup U}$, and
\[
\Trans\defeq\bigl((X_i,\pi(X_i))_{i=1}^q,\,\Uall,\,\pi|_{\Uall}\bigr).
\]
A \emph{batch collision} is a new query $u\in U$ with $\pi(u)\in D\cup U$, and
$\Ccoll=|\{u\in U:\pi(u)\in D\cup U\}|$. The pair $(D,M)$ determines this count. Querying a point
in $D$ is redundant, whereas querying a point in $\Img(P)\setminus D$ reveals a new outgoing edge.

\begin{definition}[One-batch iterate list recovery]
\label{def:task}
Fix integers $k\ge2$, $N\ge m\ge1$, and $q,b\in\Zge$. On $\CompRel_k(N,m,b)$, a learner observes
$q$ i.i.d.\ passive edges and one nonadaptive batch of at most $b$ membership queries, then outputs
$((R_1,C_1),\dots,(R_m,C_m))\in([N]^2)^m$ from $(\Trans,\omega)$. It succeeds if the starts are
pairwise distinct and $C_j=\pi^k(R_j)$ for every $j\in[m]$.
\end{definition}

\begin{remark}[Scope of the model]
\label{rem:scope}
The list size $m$ and batch budget $b$ are independent parameters. The hidden permutation is uniform,
the learner is computationally unbounded, and starts are selected after the transcript. Thus
$\CompRel_k$ measures information-theoretic certificate throughput under one passive sample and one
nonadaptive batch, rather than fixed-test-distribution generalization, computational complexity,
full-graph total-variation sampling, or multi-round interaction.
\end{remark}

\subsection{Simple prefixes and revealed walks}
\label{sub:objective}

\begin{definition}[Simple passive prefixes]
\label{def:prefix}
For $D\subseteq[N]$, $\pi\in\Perm_N$, and $a\ge1$, let
\[
\Zset_a(D)\defeq\{r\in[N]:r,\pi(r),\dots,\pi^{a-1}(r)\text{ are pairwise distinct and lie in }D\},
\]
and $Z_a(D)=|\Zset_a(D)|$.
\end{definition}

Every start in $\Zset_a(D)$ determines $\pi^a(r)=P^a(r)$ from $P$ alone. A start in $\Zset_k(D)$
needs no query, while a start in $\Zset_{k-1}(D)$ determines a query location $\pi^{k-1}(r)$ whose
answer completes the certificate.

\begin{definition}[Revealed walks and passive walk counts]
\label{def:revealed-walks}
Let $M$ be a partial injection on $[N]$ and fix $k\ge2$. A start $r$ has a revealed $k$-walk in $M$
if $M^i(r)$ is defined and belongs to $\Dom(M)$ for every $i\in\{0,\dots,k-1\}$. Write $W_k(M)$ for
the number of such starts. For $D\subseteq[N]$ and $a\ge1$, define
\[
\Hset_a(D)\defeq\{r\in[N]:r,\pi(r),\dots,\pi^{a-1}(r)\in D\},
\qquad H_a(D)\defeq|\Hset_a(D)|.
\]
\end{definition}

Thus $W_k(P)=H_k(D)$, and the revealed-walk condition uses only $M$. Always
$\Zset_a(D)\subseteq\Hset_a(D)$; their difference consists of starts on cycles of length less than
$a$ whose entire cycle lies in $D$.

\section{Main Results}
\label{sec:results}

Our first theorem determines the sharp sample threshold on the polynomial mixed profile. Here $h$ is
the query-free deficit, and $m$ is the total number of $(k-1)$-prefixes required for free and paid
certificates.

\begin{theorem}[Sharp two-branch threshold]
\label{thm:two-branch}
Suppose the following conditions hold:
\begin{itemize}
\item Fix an integer $k\ge2$ and real exponents $0<\delta<\gamma<1$.
\item For each integer $N\ge1$, set $m=\lfloor N^\gamma\rfloor$, $h=\lfloor N^\delta\rfloor$, and
$b=m-h$.
\item Define
\begin{align}
Q_{k-1}(N)&\defeq(mN^{k-2})^{1/(k-1)},\qquad
Q_k(N)\defeq(hN^{k-1})^{1/k},\notag\\
Q(N)&\defeq\max\{Q_{k-1}(N),Q_k(N)\}.\label{eq:two-scales}
\end{align}
\end{itemize}
Then for every fixed $\varepsilon\in(0,1)$, the following hold on $\CompRel_k(N,m,b)$ as
$N\to\infty$:
\begin{itemize}
\item \textbf{Part 1 (Lower bound).} Uniformly over integers $0\le q\le(1-\varepsilon)Q(N)$, every
possibly randomized, computationally unbounded learner using $q$ passive examples and at most $b$
queries in one nonadaptive batch succeeds with probability $o(1)$.
\item \textbf{Part 2 (Upper bound).} Uniformly over integers $q\ge(1+\varepsilon)Q(N)$, an explicit
learner using at most $q$ passive examples and at most $b$ queries in one nonadaptive batch succeeds
with probability $1-o(1)$.
\end{itemize}
\end{theorem}

The exponents in \eqref{eq:two-scales} are
$\alpha=(\gamma+k-2)/(k-1)$ and $\beta=(\delta+k-1)/k$, both in $(0,1)$. Thus
$Q(N)=\Theta(N^{\max(\alpha,\beta)})=o(N)$. The critical window $q=(1+o(1))Q(N)$ is outside the
statement.

\begin{corollary}[Sharp passive-only baseline]
\label{cor:passive-baseline}
Fix $k\ge2$, $\gamma\in(0,1)$, and $\varepsilon\in(0,1)$. Set $m=\lfloor N^\gamma\rfloor$, $b=0$,
and $\Qpass(N)=(mN^{k-1})^{1/k}$. Then on $\CompRel_k(N,m,0)$, every learner succeeds with
probability $o(1)$ uniformly for $0\le q\le(1-\varepsilon)\Qpass(N)$, and an explicit learner
succeeds with probability $1-o(1)$ uniformly for $q\ge(1+\varepsilon)\Qpass(N)$.
\end{corollary}

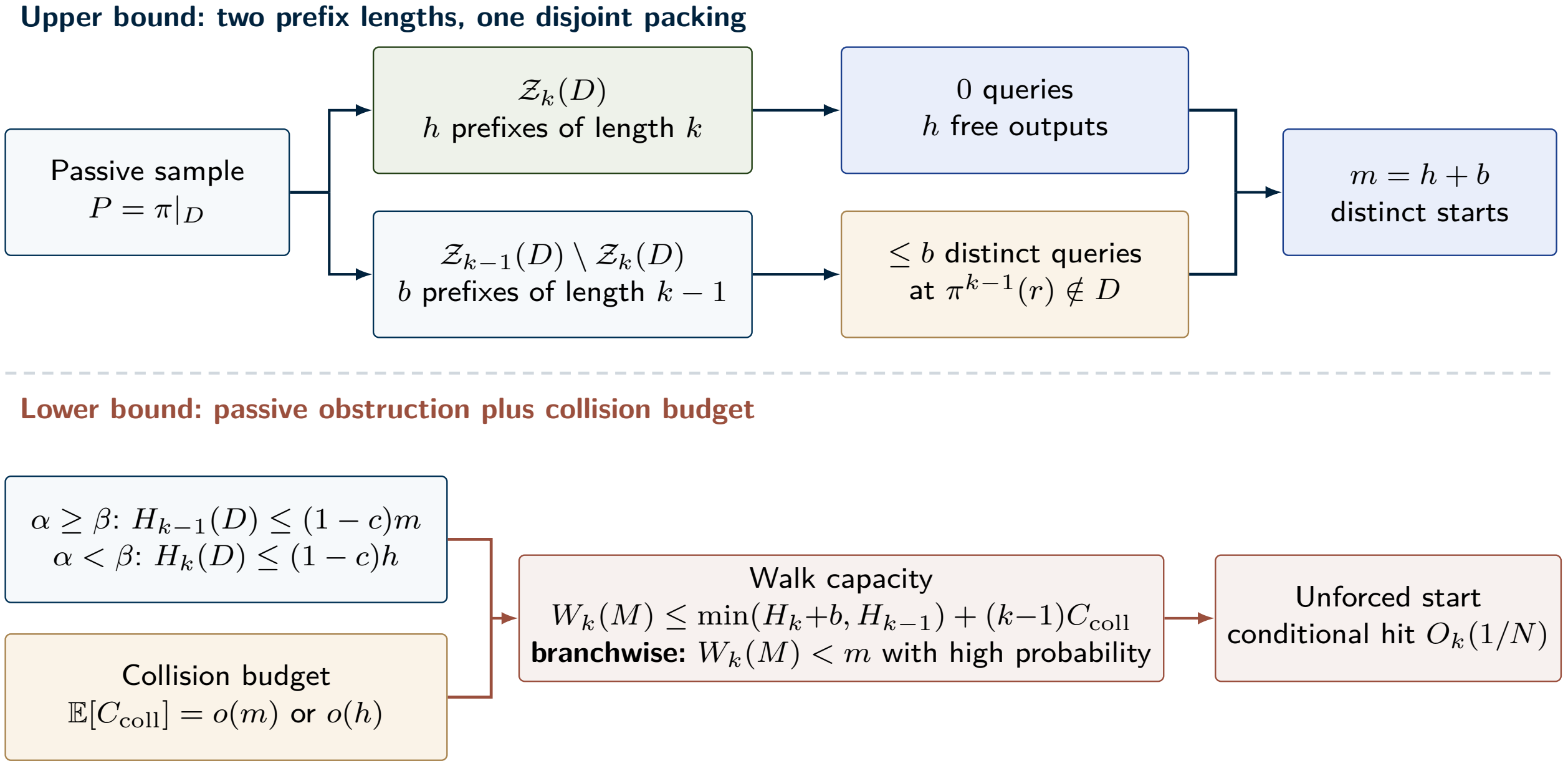
\begin{figure}[t]
\centering
\resizebox{0.98\textwidth}{!}{\begin{tikzpicture}[
  font=\scriptsize\sffamily,
  box/.style={draw=PaperAccent, rounded corners=1.5pt, fill=PaperAccentPale,
    align=center, inner sep=3.2pt, minimum height=12mm, font=\scriptsize\sffamily},
  defbox/.style={box, draw=PaperDefinition, fill=PaperDefinitionLight},
  navbox/.style={box, draw=PaperNavigation, fill=PaperNavigationLight},
  goldbox/.style={box, draw=PaperGold!85!black, fill=PaperGoldLight},
  redbox/.style={box, draw=PaperDraft, fill=PaperDraft!8},
  arr/.style={-{Latex[length=1.6mm]}, line width=0.7pt, draw=PaperAccentDark},
  redarr/.style={-{Latex[length=1.6mm]}, line width=0.7pt, draw=PaperDraft}
]

\node[font=\scriptsize\sffamily\bfseries, text=PaperAccentDark, anchor=west] at (0.10, 5.25)
  {Upper bound: two prefix lengths, one disjoint packing};
\node[box, minimum width=2.70cm] (pass) at (1.45, 3.60) {Passive sample\\ $P=\pi|_D$};
\node[defbox, minimum width=3.60cm] (zk) at (5.40, 4.38)
  {$\Zset_k(D)$\\ $h$ prefixes of length $k$};
\node[box, minimum width=3.60cm] (zk1) at (5.40, 2.82)
  {$\Zset_{k-1}(D)\setminus\Zset_k(D)$\\ $b$ prefixes of length $k-1$};
\node[navbox, minimum width=3.30cm] (free) at (9.70, 4.38) {$0$ queries\\ $h$ free outputs};
\node[goldbox, minimum width=3.30cm] (pay) at (9.70, 2.82)
  {$\le b$ distinct queries\\ at $\pi^{k-1}(r)\notin D$};
\node[navbox, minimum width=2.60cm] (out) at (13.55, 3.60) {$m=h+b$\\ distinct starts};
\coordinate (split) at (3.18, 3.60);
\coordinate (merge) at (11.80, 3.60);
\draw[line width=0.8pt, draw=PaperAccentDark] (pass.east) -- (split);
\draw[arr] (split) |- (zk.west);
\draw[arr] (split) |- (zk1.west);
\draw[arr] (zk.east) -- (free.west);
\draw[arr] (zk1.east) -- (pay.west);
\draw[line width=0.8pt, draw=PaperAccentDark] (free.east) -| (merge);
\draw[line width=0.8pt, draw=PaperAccentDark] (pay.east) -| (merge);
\draw[arr] (merge) -- (out.west);

\draw[PaperRule, densely dashed, line width=0.75pt] (0.10, 1.88) -- (14.85, 1.88);
\node[font=\scriptsize\sffamily\bfseries, text=PaperDraft, anchor=west] at (0.10, 1.52)
  {Lower bound: passive obstruction plus collision budget};
\node[box, minimum width=4.20cm] (hsmall) at (2.20, 0.30)
  {$\alpha\ge\beta$: $H_{k-1}(D)\le(1-c)m$\\ $\alpha<\beta$: $H_k(D)\le(1-c)h$};
\node[goldbox, minimum width=4.20cm] (csmall) at (2.20, -1.20)
  {Collision budget\\ $\E[\Ccoll]=o(m)$ or $o(h)$};
\node[redbox, minimum width=5.60cm] (cap) at (8.05, -0.45)
  {Walk capacity\\ $W_k(M)\le\min(H_k{+}b,H_{k-1})+(k{-}1)\Ccoll$\\ \textbf{branchwise:} $W_k(M)<m$ with high probability};
\node[redbox, minimum width=3.20cm] (post) at (13.25, -0.45)
  {Unforced start\\ conditional hit $O_k(1/N)$};
\coordinate (lmerge) at (4.72, -0.45);
\draw[line width=0.8pt, draw=PaperDraft] (hsmall.east) -| (lmerge);
\draw[line width=0.8pt, draw=PaperDraft] (csmall.east) -| (lmerge);
\draw[redarr] (lmerge) -- (cap.west);
\draw[redarr] (cap.east) -- (post.west);
\end{tikzpicture}}
\caption{Certificate accounting for \Cref{thm:two-branch}. In the upper panel, Corn boxes denote
passive $k$-prefixes, Prussian Blue boxes denote $(k-1)$-prefixes, and Champagne Gold denotes paid
queries; these combine into $m$ distinct starts. In the lower panel, passive obstruction and the
Champagne Gold collision budget feed the warm-red capacity inequality of \Cref{lem:collision-capacity},
which leaves an unforced start with conditional hit probability $O_k(1/N)$.}
\label{fig:allocation}
\end{figure}

\paragraph{Proof architecture.}
Above $(1+\varepsilon)Q(N)$, domain concentration preserves the margin and
\Cref{lem:prefix-moments} supplies both prefix quotas. The learner packs $h$ free starts from
$\Zset_k(D)$ and $b$ further starts from $\Zset_{k-1}(D)\setminus\Zset_k(D)$, querying
$P^{k-1}(r)$ only when it is not already in $D$.

Below $(1-\varepsilon)Q(N)$, monotone coupling and Chebyshev concentration bound the relevant
passive walk count by $(1-c)m$ or $(1-c)h$. \Cref{lem:collision-capacity} gives
\[
W_k(M)\le\min\bigl(H_k(D)+b,\,H_{k-1}(D)\bigr)+(k-1)\Ccoll.
\]
Since $m-b=h$, either branch leaves $W_k(M)<m$ with high probability. Fixing the seed before
conditioning on the transcript then permits \Cref{lem:last-fresh} to bound the conditional success
probability by $O_k(1/N)$.

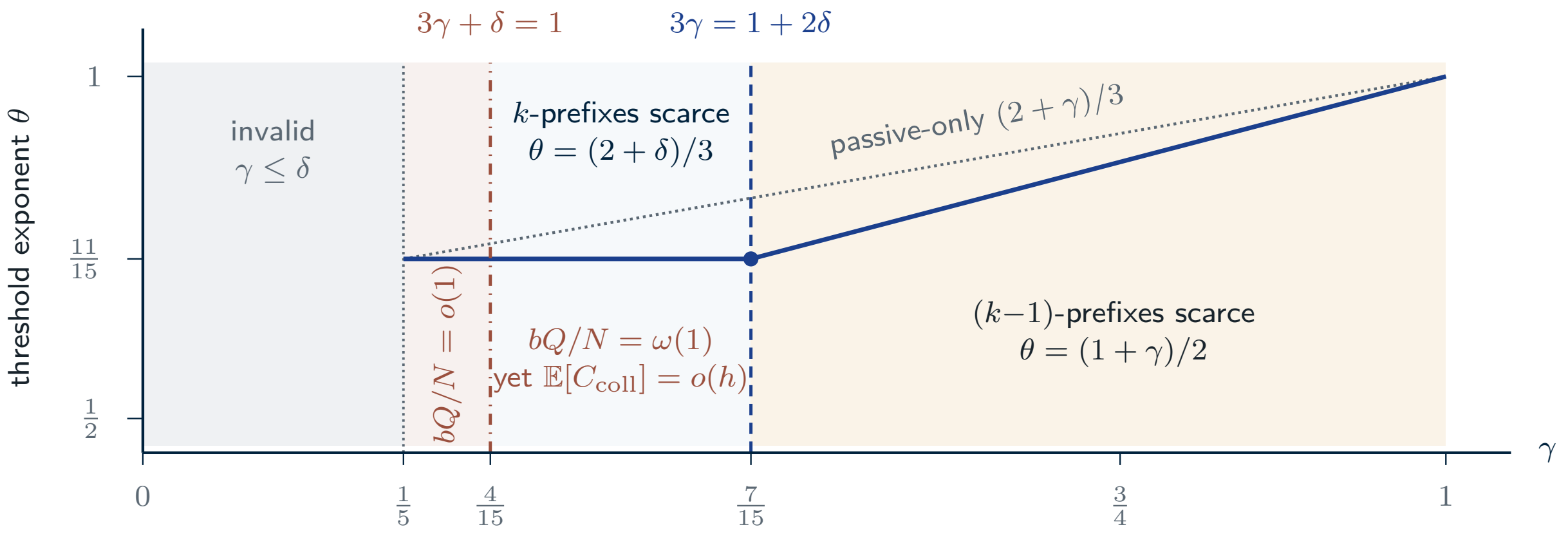
\begin{figure}[t]
\centering
\resizebox{0.96\textwidth}{!}{\begin{tikzpicture}[
  x=12.2cm, y=6.4cm,
  font=\scriptsize\sffamily,
  axis/.style={line width=0.75pt, draw=PaperAccentDark},
  kink/.style={line width=1.2pt, draw=PaperNavigation},
  passive/.style={line width=0.75pt, draw=PaperMutedInk, densely dotted}
]
  \fill[PaperRule!35] (0.0,0.46) rectangle (0.20,1.02);
  \fill[PaperDraft!8] (0.20,0.46) rectangle (0.2666667,1.02);
  \fill[PaperAccentPale] (0.2666667,0.46) rectangle (0.4666667,1.02);
  \fill[PaperGoldLight] (0.4666667,0.46) rectangle (1.00,1.02);
  \draw[densely dotted, line width=0.65pt, draw=PaperMutedInk] (0.20,0.45) -- (0.20,1.02);
  \node[align=center, font=\scriptsize\sffamily, text=PaperMutedInk] at (0.10,0.89)
    {invalid\\ $\gamma\le\delta$};

  \draw[passive] (0.20,0.7333333) -- (1.00,1.00);
  \node[rotate=10, font=\scriptsize\sffamily, text=PaperMutedInk] at (0.64,0.935)
    {passive-only $(2+\gamma)/3$};
  \draw[kink] (0.20,0.7333333) -- (0.4666667,0.7333333) -- (1.00,1.00);
  \fill[PaperNavigation] (0.4666667,0.7333333) circle (2.0pt);

  \draw[dash dot, line width=0.85pt, draw=PaperDraft] (0.2666667,0.45) -- (0.2666667,1.02);
  \node[anchor=south, font=\scriptsize\sffamily, text=PaperDraft] at (0.2666667,1.035)
    {$3\gamma+\delta=1$};
  \draw[densely dashed, line width=0.85pt, draw=PaperNavigation] (0.4666667,0.45) -- (0.4666667,1.02);
  \node[anchor=south, font=\scriptsize\sffamily, text=PaperNavigation] at (0.4666667,1.035)
    {$3\gamma=1+2\delta$};

  \draw[axis] (0,0.45) -- (1.05,0.45);
  \draw[axis] (0,0.45) -- (0,1.07);
  \node[anchor=west, font=\scriptsize\sffamily\bfseries, text=PaperAccentDark] at (1.06,0.45) {$\gamma$};
  \node[anchor=south, rotate=90, font=\scriptsize\sffamily, text=PaperInk] at (-0.072,0.75)
    {threshold exponent $\theta$};
  \foreach \g/\lab in {0/0,0.20/{\tfrac15},0.2666667/{\tfrac4{15}},0.4666667/{\tfrac7{15}},0.75/{\tfrac34},1/1} {
    \draw[PaperAccentDark] (\g,0.45) -- (\g,0.432);
    \node[anchor=north, font=\scriptsize\sffamily, text=PaperMutedInk] at (\g,0.422) {$\lab$};
  }
  \foreach \y/\lab in {0.5/{\tfrac12},0.7333333/{\tfrac{11}{15}},1/1} {
    \draw[PaperAccentDark] (0,\y) -- (-0.012,\y);
    \node[anchor=east, font=\scriptsize\sffamily, text=PaperMutedInk] at (-0.020,\y) {$\lab$};
  }
  \node[rotate=90, font=\scriptsize\sffamily, text=PaperDraft] at (0.2333,0.595) {$bQ/N=o(1)$};
  \node[align=center, font=\scriptsize\sffamily, text=PaperDraft] at (0.367,0.585)
    {$bQ/N=\omega(1)$\\ yet $\E[\Ccoll]=o(h)$};
  \node[align=center, font=\scriptsize\sffamily, text=PaperAccentDark] at (0.367,0.915)
    {$k$-prefixes scarce\\ $\theta=(2+\delta)/3$};
  \node[align=center, font=\scriptsize\sffamily, text=PaperInk] at (0.745,0.625)
    {$(k{-}1)$-prefixes scarce\\ $\theta=(1+\gamma)/2$};
\end{tikzpicture}}
\caption{Depth-$3$ running example at $\delta=1/5$, with legal domain $\gamma>1/5$. The solid
Mingjing Blue polyline is the threshold exponent $\theta(\gamma)=\max\{(2+\delta)/3,(1+\gamma)/2\}$,
and the dotted gray line is the passive-only exponent $(2+\gamma)/3$. The vertical dash-dot warm-red
line $3\gamma+\delta=1$ separates the pale-red zero-collision-union-bound strip from the pale-blue
region where that bound diverges but $\E[\Ccoll]=o(h)$. The vertical densely dashed Mingjing Blue
line $3\gamma=1+2\delta$ marks the actual exponent kink and the start of the pale-gold
$(k-1)$-prefix-scarce regime; the gray region $\gamma\le\delta$ is outside the mixed profile.}
\label{fig:phase}
\end{figure}

\paragraph{Separation from the zero-collision boundary.}
In this example, the union bound $b(q+b)/(N-q)$ crosses from $o(1)$ to $\omega(1)$ at
$3\gamma+\delta=1$, yet remains $o(h)$ throughout $(4/15,7/15)$. The threshold therefore bends only
at the supply crossover $3\gamma=1+2\delta$. Retaining $\Ccoll$ in the capacity inequality is what
separates these two lines.

\section{Proof of the Sharp Mixed-Profile Threshold}
\label{sec:proofs}

This section proves \Cref{thm:two-branch} and \Cref{cor:passive-baseline}. Throughout,
$k\ge2$ is fixed, $m,h,b,Q(N)$ are as in \eqref{eq:two-scales}, and
$\alpha=(\gamma+k-2)/(k-1)$, $\beta=(\delta+k-1)/k$.

\subsection{Domain size and prefix moments}
\label{sub:prefix-supply}

\begin{lemma}[Passive domain loss]
\label{lem:domain-reduction}
Fix integers $N\ge1$ and $q_0\ge1$. Let $X_1,\dots,X_{q_0}$ be i.i.d.\ uniform on $[N]$ and set
$D=\{X_1,\dots,X_{q_0}\}$. Then $\E[q_0-|D|]\le\binom{q_0}{2}/N$, and for every $\rho>0$,
\[
\Prob[|D|<(1-\rho)q_0]\le\frac{(q_0-1)_+}{2\rho N}.
\]
\end{lemma}

\begin{proof}
The duplicate count satisfies
\[
q_0-|D|=\sum_{i=2}^{q_0}\ind\{X_i\in\{X_1,\dots,X_{i-1}\}\}
\le\sum_{1\le j<i\le q_0}\ind\{X_i=X_j\}.
\]
Taking expectations gives the first bound. For $q_0=1$, $|D|=1$ and the event is empty. For
$q_0\ge2$, Markov's inequality applied to $q_0-|D|>\rho q_0$ gives the second bound.
\end{proof}

\begin{lemma}[Simple prefix moments]
\label{lem:prefix-moments}
Fix integers $a\ge1$, $N\ge2a$, and $2a\le d\le N$. Let $D$ be a uniform $d$-subset independent of
$\pi$. Then
\[
\lambda_a(d)\defeq\E[Z_a(D)]=(N-a+1)\frac{(d)_a}{(N)_a},
\qquad
\Var(Z_a(D))\le(4a-2)\lambda_a(d)+B_a\frac{\lambda_a(d)^2}{N},
\]
where $B_a>0$ depends only on $a$.
\end{lemma}

\begin{proof}
Let $L(r)$ be the cycle length containing $r$ and $O_a(r)=\{r,\pi(r),\dots,\pi^{a-1}(r)\}$. By
\Cref{lem:cycle-length}, $\Prob[L(r)\ge a]=(N-a+1)/N$. Conditional on this event,
$\Prob[O_a(r)\subseteq D\mid\pi]=(d)_a/(N)_a$, which gives the mean.

For the second moment, partition ordered pairs $(r,s)$. If $O_a(r)$ and $O_a(s)$ intersect, then
$s=\pi^t(r)$ for some $|t|\le a-1$, so there are at most $(2a-1)N$ such pairs. Each contributes at
most $(d)_a/(N)_a$. Since $N\ge2a$ implies $N-a+1\ge N/2$, the total contribution is at most
$(4a-2)\lambda_a(d)$. Disjoint pairs have $2a$ vertices and contribute at most
$N^2(d)_{2a}/(N)_{2a}$. Using $(d)_{2a}\le(d)_a^2$ and $(N)_{2a}=(N)_a(N-a)_a$,
\[
N^2\frac{(d)_{2a}}{(N)_{2a}}
\le\frac{N^2}{(N-a+1)^2}\frac{(N)_a}{(N-a)_a}\lambda_a(d)^2
=R_a(N)\lambda_a(d)^2.
\]
\Cref{lem:falling-factorial-ratio} gives $R_a(N)\le1+B_a/N$. Subtracting $\lambda_a(d)^2$ proves the
variance bound.
\end{proof}

\begin{lemma}[Short-cycle correction]
\label{lem:short-cycles}
Fix $1\le a\le N$. Let $D$ be independent of $\pi$ and uniform among subsets of its size. Then
\[
\E[H_a(D)-Z_a(D)\mid |D|=d]=\sum_{\ell=1}^{a-1}\frac{(d)_\ell}{(N)_\ell}.
\]
If $|D|\le q$ almost surely with $q=o(N)$, then
$\Prob[H_a(D)\ne Z_a(D)]\le(a-1)q/N=o(1)$.
\end{lemma}

\begin{proof}
A start belongs to $\Hset_a(D)\setminus\Zset_a(D)$ exactly when its cycle length is $\ell<a$ and its
entire $\ell$-cycle lies in $D$. This occurs with probability $(d)_\ell/(N)_\ell$ conditional on
$L(r)=\ell$. Summing over $r$ and $\ell$ gives the identity. Each term is at most $q/N$, and the
nonnegative integer correction is nonzero with probability at most its expectation.
\end{proof}

\subsection{Posterior uniformity and walk capacity}
\label{sub:posterior-capacity}

\begin{lemma}[Uniform completions]
\label{lem:posterior-completion}
For almost every $\omega=w$, the batch and output maps are deterministic functions of $(X,P)$ and
$\Trans$, respectively. Conditional on $(X,P,\omega=w)$, $\pi$ is uniform over completions of $P$;
conditional on $(\Trans,\omega=w)$, $\pi$ is uniform over completions of $M$.
\end{lemma}

\begin{proof}
Independence of $\omega$ and $(\pi,X)$ preserves their joint law after conditioning on $\omega=w$.
For fixed $(x,p)$, every completion $\sigma$ of $p$ has joint probability
$\Prob[\pi=\sigma,X=x\mid\omega=w]=(N!N^q)^{-1}$, independent of $\sigma$. The batch is then a
deterministic function of $(x,p)$, and compatible batch answers extend the same constancy to every
completion of $M$.
\end{proof}

\begin{lemma}[Batch collision expectation and deterministic walk capacity]
\label{lem:collision-capacity}
Fix $k\ge2$ and $q,b\in\Zge$ with $q+b<N$. Then
\[
\E[\Ccoll]\le\frac{b(q+b)}{N-q},
\qquad
W_k(M)\le\min\bigl(H_k(D)+b,\,H_{k-1}(D)\bigr)+(k-1)\Ccoll
\]
deterministically.
\end{lemma}

\begin{proof}
Conditional on $(X,P,\omega)$, each $u\in U$ has image uniform on the $N-|D|$ values outside
$\Img(P)$, and at most $q+b$ of those values lie in $D\cup U$. Linearity gives the expectation
bound.

For the capacity bound, let $\mathcal W$ be the revealed starts, write $v_i=\pi^i(r)$, and let
$J(r)$ be the least index with $v_J\in U$, or $\infty$ if none exists.
\begin{itemize}
\item If $J=\infty$, then $r\in\Hset_k(D)\subseteq\Hset_{k-1}(D)$, contributing at most $H_k(D)$.
\item If $J=k-1$, then $r\in\Hset_{k-1}(D)$, and $r\mapsto\pi^{k-1}(r)$ injects these starts into
$U$, contributing at most $b$.
\item If $J=j\le k-2$, then $v_j\in U$ and $v_{j+1}\in D\cup U$, so $v_j$ is a collision. For each
$j$, $r\mapsto\pi^j(r)$ is injective, giving at most $\Ccoll$ starts per $j$ and at most
$(k-1)\Ccoll$ in total.
\end{itemize}
The first and second classes are disjoint subsets of $\Hset_{k-1}(D)$, which yields the second bound
in the minimum.
\end{proof}

\begin{figure}[t]
\centering
\resizebox{0.98\textwidth}{!}{\begin{tikzpicture}[
  font=\scriptsize\sffamily,
  dnode/.style={circle, draw=PaperDefinition, fill=PaperDefinitionLight, inner sep=1.8pt, minimum size=6mm},
  unode/.style={circle, draw=PaperGold!85!black, fill=PaperGoldLight, inner sep=1.8pt, minimum size=6mm},
  cnode/.style={circle, draw=PaperDraft, fill=PaperDraft!8, line width=0.9pt, inner sep=1.8pt, minimum size=6mm},
  endnode/.style={circle, draw=PaperAccentDark, densely dashed, fill=PaperAccentPale, inner sep=1.8pt, minimum size=6mm},
  pedge/.style={-{Latex[length=1.5mm]}, line width=0.75pt, draw=PaperDefinition},
  qedge/.style={-{Latex[length=1.5mm]}, line width=0.9pt, densely dashed, draw=PaperNavigation},
  cedge/.style={-{Latex[length=1.5mm]}, line width=1.1pt, draw=PaperDraft},
  nedge/.style={-{Latex[length=1.5mm]}, line width=0.7pt, draw=PaperAccentDark},
  sumbox/.style={draw, rounded corners=1.5pt, align=left, inner sep=3pt, minimum width=4.7cm, font=\scriptsize\sffamily}
]
\node[font=\scriptsize\sffamily\bfseries, text=PaperAccentDark, anchor=west] at (0,4.10)
  {(a) Walk-class partition (shown for $k=3$)};
\node[dnode, minimum size=3.4mm] at (10.35,4.10) {};
\node[anchor=west, text=PaperMutedInk] at (10.56,4.10) {$\in D$};
\node[unode, minimum size=3.4mm] at (11.55,4.10) {};
\node[anchor=west, text=PaperMutedInk] at (11.76,4.10) {$\in U$};
\node[cnode, minimum size=3.4mm] at (12.75,4.10) {};
\node[anchor=west, text=PaperMutedInk] at (12.96,4.10) {hit in $D\cup U$};

\node[anchor=west, text=PaperDefinition] at (0,3.15) {\textbf{Class 1}: $J=\infty$};
\node[dnode] (a10) at (3.20,3.15) {$v_0$};
\node[dnode] (a11) at (5.10,3.15) {$v_1$};
\node[dnode] (a12) at (7.00,3.15) {$v_2$};
\node[endnode] (a13) at (8.90,3.15) {$v_3$};
\draw[pedge] (a10) -- (a11); \draw[pedge] (a11) -- (a12); \draw[pedge] (a12) -- (a13);
\node[sumbox, draw=PaperDefinition, fill=PaperDefinitionLight] at (12.60,3.15)
  {$\mathcal W_1\subseteq\Hset_3(D)\subseteq\Hset_2(D)$\\ $|\mathcal W_1|\le H_3(D)$};

\node[anchor=west, text=PaperNavigation] at (0,1.95) {\textbf{Class 2}: $J=2$};
\node[dnode] (a20) at (3.20,1.95) {$v_0$};
\node[dnode] (a21) at (5.10,1.95) {$v_1$};
\node[unode] (a22) at (7.00,1.95) {$v_2$};
\node[endnode] (a23) at (8.90,1.95) {$v_3$};
\draw[pedge] (a20) -- (a21); \draw[pedge] (a21) -- (a22); \draw[qedge] (a22) -- (a23);
\node[sumbox, draw=PaperNavigation, fill=PaperNavigationLight] at (12.60,1.95)
  {$\mathcal W_2\subseteq\Hset_2(D)$\\ $r\mapsto v_2$ injective, $|\mathcal W_2|\le b$};

\node[anchor=west, text=PaperDraft] at (0,0.75) {\textbf{Class 3}: $J\le1$};
\node[dnode] (a30) at (3.20,0.75) {$v_0$};
\node[unode] (a31) at (5.10,0.75) {$v_1$};
\node[cnode] (a32) at (7.00,0.75) {$v_2$};
\node[endnode] (a33) at (8.90,0.75) {$v_3$};
\draw[pedge] (a30) -- (a31); \draw[cedge] (a31) -- (a32); \draw[nedge] (a32) -- (a33);
\node[sumbox, draw=PaperDraft, fill=PaperDraft!8] at (12.60,0.75)
  {each $j$: $r\mapsto v_j$ injective\\ $|\mathcal W_3|\le(k{-}1)\Ccoll$};

\draw[PaperRule, densely dashed, line width=0.75pt] (0,-0.05) -- (15.0,-0.05);
\node[font=\scriptsize\sffamily\bfseries, text=PaperDraft, anchor=west] at (0,-0.42)
  {(b) Nonterminal collision synergy ($k=2$)};
\node[dnode] (b0) at (1.00,-1.85) {$r$};
\node[unode] (b1) at (3.45,-1.85) {$u$};
\node[cnode, fill=PaperDefinitionLight] (b2) at (5.90,-1.85) {$w$};
\node[endnode] (b3) at (8.35,-1.85) {$P(w)$};
\draw[pedge] (b0) -- (b1); \draw[cedge] (b1) -- (b2); \draw[pedge] (b2) -- (b3);
\draw[line width=0.8pt, draw=PaperNavigation]
  (1.00,-1.32) -- (1.00,-1.16) -- (5.90,-1.16) -- (5.90,-1.32);
\node[anchor=south, font=\scriptsize\sffamily, text=PaperNavigation] at (3.45,-1.13)
  {terminal start $(r,w)$};
\draw[line width=0.8pt, draw=PaperGold!85!black]
  (3.45,-2.38) -- (3.45,-2.54) -- (8.35,-2.54) -- (8.35,-2.38);
\node[anchor=north, font=\scriptsize\sffamily, text=PaperGold!65!black] at (5.90,-2.57)
  {nonterminal start $(u,P(w))$};
\node[sumbox, draw=PaperGold!85!black, fill=PaperGoldLight, minimum height=18mm] at (12.60,-1.85)
  {\textbf{One query, two starts}\\ $S_{\mathrm{term}}\subseteq D\setminus\Hset_2(D)$ and
   $S_{\mathrm{coll}}\subseteq U$ are disjoint.};
\end{tikzpicture}}
\caption{Walk-class decomposition and nonterminal collision synergy. In panel (a), Corn circles and
solid edges are passive, Champagne Gold circles are new queries, and warm-red rings mark collisions;
the three classes give the bounds in \Cref{lem:collision-capacity}. In panel (b), one query at
$u=\pi(r)$ landing at $w\in D$ certifies the terminal start $r$ and the disjoint nonterminal start
$u$, illustrating \Cref{prop:k2-counterexample}.}
\label{fig:walk-classes}
\end{figure}
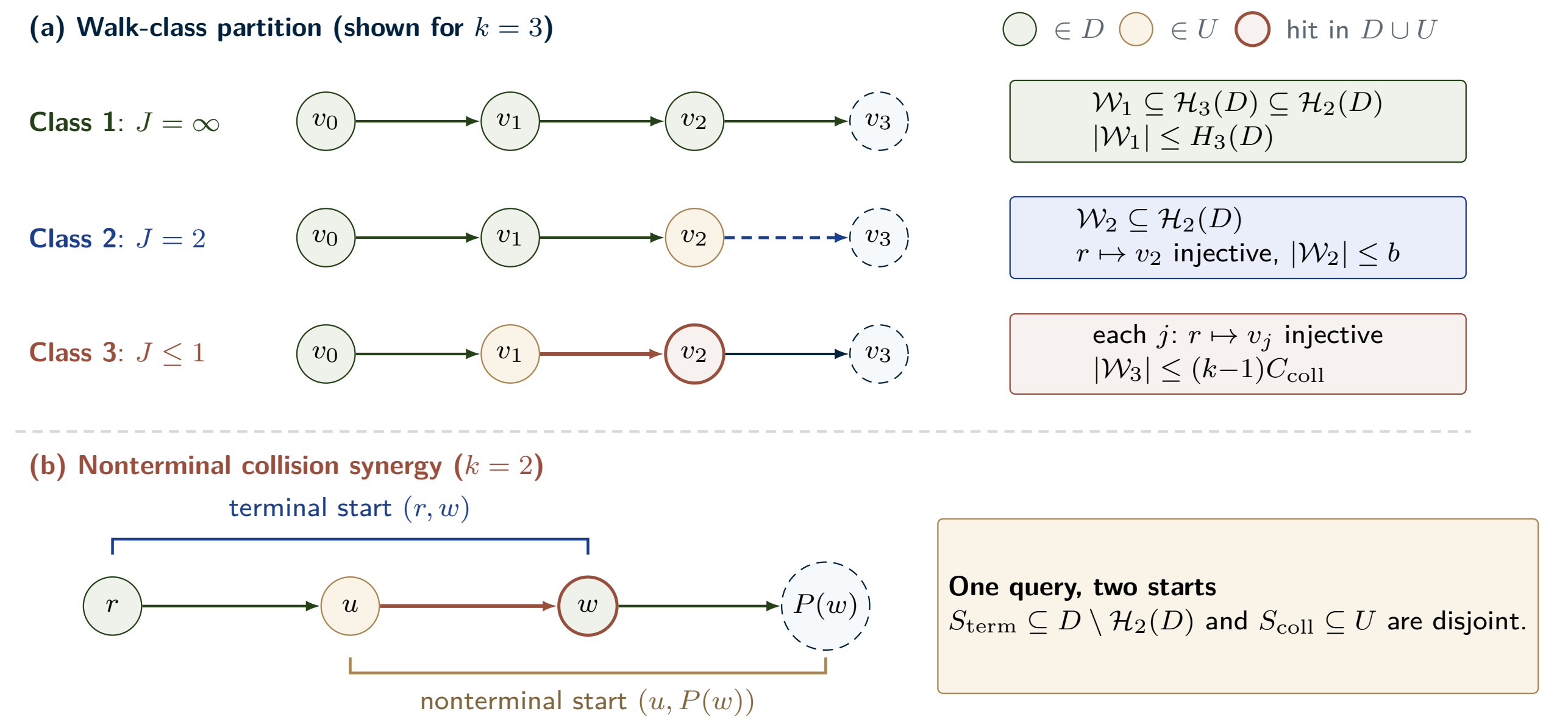

\begin{lemma}[Last-fresh-edge posterior bound]
\label{lem:last-fresh}
Let $M$ be deterministic with $|\Dom(M)|=d$, $N>d+k$, and let $\pi$ be a uniform completion. If $r$
has no revealed $k$-walk in $M$, then for every $c\in[N]$,
\[
\Prob[\pi^k(r)=c]\le\sum_{j=0}^{k-1}\frac{k-j}{N-d-j}
\le\frac{k(k+1)}{2(N-d-k+1)}.
\]
\end{lemma}

\begin{proof}
Set $Y_0=r$ and $M_0=M$. Sequentially set $Y_{j+1}=\pi(Y_j)$, let
$F_j=\{Y_j\notin\Dom(M_j)\}$, and add the edge $(Y_j,Y_{j+1})$ to $M_j$ only when $F_j$ occurs.
For every realizable tuple $(y_0,\dots,y_j,m_j)$, the path event is exactly the uniform-completion
event $m_j\subseteq\pi$: starting at $r$ and reading $m_j$ reproduces the path, and every added edge
belongs to $\pi$. Thus on $F_j$, $Y_{j+1}$ is uniform on $[N]\setminus\Img(M_j)$, whose size is at
least $N-d-j$.

At least one of $F_0,\dots,F_{k-1}$ occurs. Let $J$ be the last such index, and set
$s=k-1-j$. Given $J=j$ and a fresh image $v$, the remaining $s$ steps are non-fresh. Either they
never revisit $Y_j$, forcing $M_j^s(v)=c$, or they first return at some $t\in\{0,\dots,s-1\}$,
forcing $M_j^t(v)=Y_j$. Each alternative has at most one solution by injectivity, so at most
$k-j$ values of $v$ are admissible. Summing these bounds over $j$ proves the lemma.
\end{proof}

\subsection{Proof of the sharp upper bound}
\label{sub:upper-bound}

\begin{proof}[Proof of \Cref{thm:two-branch}, Part~2]
Set $q_0=\lceil(1+\varepsilon)Q(N)\rceil$ and discard samples beyond $q_0$. Choose $\rho>0$ with
$(1-\rho)(1+\varepsilon)>1$, define $1+2c=((1-\rho)(1+\varepsilon))^{k-1}$, and set
$\tau=c/(4(1+c))$. By \Cref{lem:domain-reduction}, $|D|\ge(1-\rho)q_0$ with probability $1-o(1)$.
Condition on $|D|=d$ in this range. Uniformly,
$\lambda_a(d)=(1-o(1))d^a/N^{a-1}$, so $\lambda_{k-1}(d)\ge(1+c)m$ and
$\lambda_k(d)\ge(1+c)h$ for large $N$.

For $Z_k$, the event $Z_k<h$ implies a deviation of at least $c\lambda_k/(1+c)$, hence Chebyshev
gives failure probability at most
\[
\frac{(1+c)^2}{c^2}\left(\frac{4k-2}{(1+c)h}+\frac{B_k}{N}\right)=o(1).
\]
Similarly, $Z_{k-1}\ge(1-\tau)\lambda_{k-1}$ with probability $1-o(1)$ (deterministically when
$k=2$). The exact mean ratio satisfies
\[
\frac{\lambda_k(d)}{\lambda_{k-1}(d)}=\frac{d-k+1}{N-k+2}\le\frac{q_0}{N-k+2}=o(1),
\]
so Markov gives $Z_k\le\tau\lambda_{k-1}$ with probability $1-o(1)$. On the intersection,
\[
Z_{k-1}-Z_k\ge(1-2\tau)(1+c)m=(1+c/2)m\ge m\ge b.
\]
Thus the packing event holds with probability $1-o(1)$. The deterministic procedure in
\Cref{alg:mixed-packing} selects $h$ free starts and $b$ paid starts, uses at most $b$ distinct
queries, and succeeds on that event by \Cref{prop:packing-deterministic}.
\end{proof}

\subsection{Proof of the sharp lower bound and passive baseline}
\label{sub:lower-bound}

\begin{proof}[Proof of \Cref{thm:two-branch}, Part~1]
Set $q_{\max}=\lfloor(1-\varepsilon)Q(N)\rfloor$ and fix any $0\le q\le q_{\max}$. Conditional on
$|D|=d$, augment $D$ by uniformly chosen outside points to obtain a uniform $q_{\max}$-subset
$\widetilde D\supseteq D$ independent of $\pi$. Monotonicity gives $H_a(D)\le H_a(\widetilde D)$.
By \Cref{lem:short-cycles}, $H_a(\widetilde D)=Z_a(\widetilde D)$ with probability $1-o(1)$, and
Chebyshev gives, for every fixed $\tau>0$,
\[
\Prob[Z_a(\widetilde D)>(1+\tau)\lambda_a(q_{\max})]
\le\frac{4a-2}{\tau^2\lambda_a(q_{\max})}+\frac{B_a}{\tau^2N}=o(1),
\]
with the case $a=1$ deterministic. These bounds depend only on $q_{\max}$, so they are uniform in
$q$.

If $\alpha\ge\beta$, then $Q/Q_{k-1}\to1$ and
$\lambda_{k-1}(q_{\max})\le(1-\varepsilon+o(1))^{k-1}m$. Thus $H_{k-1}(D)\le(1-c)m$ with
probability $1-o(1)$ for some fixed $c>0$. Also $b\le m$ and $q_{\max}+b=o(N)$ imply
$\E[\Ccoll]=o(m)$, so $W_k(M)<m$ with probability $1-o(1)$ by the second capacity bound.

If $\alpha<\beta$, then $Q/Q_k\to1$ and $H_k(D)\le(1-c)h$ with probability $1-o(1)$. The condition
$\alpha<\beta$ is $k\gamma<(k-1)\delta+1$, which gives $bq_{\max}/(Nh)=o(1)$. It also gives
$b^2/(Nh)=o(1)$: for $k=2$ this is $2\gamma<1+\delta$, and for $k\ge3$ it follows from
$((k-1)\delta+1)/k<(1+\delta)/2$. Hence $\E[\Ccoll]=o(h)$, and the first capacity bound gives
$W_k(M)<m$ with probability $1-o(1)$.

Now fix $\omega=w$ before conditioning on $\Trans$. On $\mathcal B=\{W_k(M)<m\}$, either duplicate
starts cause failure or some first reported start has no revealed walk. \Cref{lem:last-fresh} bounds
its conditional success probability by $O_k(1/(N-q_{\max}-b))=o(1)$. Averaging over $\omega$ and
$\Trans$ proves the lower bound uniformly in $q$.
\end{proof}

\begin{proof}[Proof of \Cref{cor:passive-baseline}]
Set $b=0$ and $\Qpass=(mN^{k-1})^{1/k}$. The upper argument with
$q_0=\lceil(1+\varepsilon)\Qpass\rceil$ gives $Z_k(D)\ge m$ with probability $1-o(1)$, and the
packing procedure outputs $(r,P^k(r))$. For the lower bound, couple $D$ upward to a uniform
$q_{\max}$-subset as above to obtain $H_k(D)<m$ with probability $1-o(1)$. Since $U=\emptyset$,
$W_k(M)=H_k(D)$. Fix the seed before conditioning on the transcript and apply \Cref{lem:last-fresh}.
\end{proof}

\section{Independent Query Budgets}
\label{sec:decoupled}

We now allow $m_N$ and $b_N$ to vary independently. When $b_N=O(m_N)$ and the passive quota is zero
or a constant fraction of $m_N$, the sharp threshold persists. In the thin boundary, nonterminal
collisions can instead supply a constant fraction of the deficit.

\begin{theorem}[Sharp threshold outside the thin mixed-profile boundary]
\label{thm:decoupled}
Suppose the following conditions hold:
\begin{itemize}
\item Fix $k\ge2$. Let $m_N,b_N$ be integer sequences with $m_N\to\infty$, $m_N=o(N)$, and
$b_N=O(m_N)$.
\item Set $\bact=\min(m_N,b_N)$, $h_N=(m_N-b_N)_+$, and
\begin{equation}\label{eq:decoupled-scales}
A_N=(m_NN^{k-2})^{1/(k-1)},\qquad
B_N=(h_NN^{k-1})^{1/k},\qquad Q(m_N,b_N)=\max\{A_N,B_N\},
\end{equation}
with $B_N=0$ when $h_N=0$.
\item For some constant $\eta>0$, for all sufficiently large $N$, either $b_N\ge m_N$, or $b_N<m_N$
and $h_N\ge\eta m_N$.
\end{itemize}
Then for every fixed $\varepsilon\in(0,1)$, every learner succeeds with probability $o(1)$ uniformly
for $0\le q\le(1-\varepsilon)Q(m_N,b_N)$, and an explicit learner succeeds with probability $1-o(1)$
uniformly for $q\ge(1+\varepsilon)Q(m_N,b_N)$.
\end{theorem}

\begin{proof}
Since $m_N=o(N)$ and $b_N=O(m_N)$, $Q(m_N,b_N)=o(N)$ and $q+b_N=o(N)$ on the stated ranges.
For the upper bound, take $q_0=\lceil(1+\varepsilon)Q(m_N,b_N)\rceil$ and choose $\rho,c,\tau$ as
in \Cref{sub:upper-bound}. On $|D|\ge(1-\rho)q_0$, prefix concentration gives
$Z_k\ge h_N$ (trivially if $h_N=0$) and
$Z_{k-1}-Z_k\ge(1-2\tau)(1+c)m_N\ge\bact$ with probability $1-o(1)$. The packing procedure then
uses at most $\bact\le b_N$ queries and outputs $m_N$ pairs.

For the lower bound, set $q_{\max}=\lfloor(1-\varepsilon)Q(m_N,b_N)\rfloor$. The collision
expectation is $o(m_N)$. If $A_N\ge B_N$, then $Q/A_N\to1$ and
$H_{k-1}(D)\le(1-c)m_N$ with probability $1-o(1)$, so the second capacity bound gives
$W_k(M)<m_N$. If $B_N>A_N$, then $h_N>0$, hence $h_N\ge\eta m_N$, so
$\E[\Ccoll]=o(h_N)$ and $H_k(D)\le(1-c)h_N$ with probability $1-o(1)$. The first capacity bound
gives $W_k(M)\le b_N+(1-c)h_N+o(h_N)<m_N$. These estimates hold pointwise in $N$, so oscillation
between the two branches causes no uniformity problem. The posterior step is identical to
\Cref{sub:lower-bound}.
\end{proof}

The excluded thin regime is $h_N=o(m_N)$ with comparable $A_N,B_N$. There, $\E[\Ccoll]$ can be
$o(m_N)$ yet $\Theta(h_N)$, allowing nonterminal collisions to beat the naive two-scale formula.

\begin{proposition}[Nonterminal collisions beat $Q(m,b)$ inside the thin boundary]
\label{prop:k2-counterexample}
Fix $k=2$. Set $m=\lfloor N^{3/4}\rfloor$, $h=\lfloor4m^2/N\rfloor$, and $b=m-h$. Then
$Q(m,b)=(2+o(1))m$, and at $q=\lfloor1.8m\rfloor=(0.9+o(1))Q(m,b)$, an explicit learner on
$\CompRel_2(N,m,b)$ succeeds with probability $1-o(1)$.
\end{proposition}

\begin{proof}
We have $h=(4+o(1))N^{1/2}$, $b=(1-o(1))m$, and $Q(m,b)=(hN)^{1/2}=(2+o(1))m$.
By \Cref{lem:domain-reduction}, $|D|=(1.8-o(1))m$ with probability $1-o(1)$. Conditional on
$|D|=d$ in this range, \Cref{lem:prefix-moments} gives
\[
\lambda_2(d)=\frac{d(d-1)}N=(0.81+o(1))h,
\qquad \Var(Z_2(D))=O(h).
\]
Together with \Cref{lem:short-cycles}, this gives $H_2(D)=(0.81+o(1))h$ and
$|D\setminus\Hset_2(D)|>b$ with probability $1-o(1)$.

On this event, choose $b$ distinct starts $r_i\in D\setminus\Hset_2(D)$ and query
$u_i=\pi(r_i)$. These $u_i$ are distinct and lie outside $D$. Three disjoint start sets are now
certified: $\Hset_2(D)$ passively, $\{r_i\}$ terminally, and
$S_{\mathrm{coll}}=\{u_i:\pi(u_i)\in D\}$ nonterminally.

Conditional on $(X,P)$, $(\pi(u_1),\dots,\pi(u_b))$ is a uniform ordered sample without replacement
from $V=[N]\setminus\Img(P)$, of size $M_0=N-|D|$. The number of targets in $D$ is
$K_0=|D|-H_2(D)$, because $\pi$ bijects $\Hset_2(D)$ with $D\cap\Img(P)$. Hence
$|S_{\mathrm{coll}}|$ is hypergeometric with mean $\mu_{\mathrm{coll}}=bK_0/M_0$ and variance at
most $\mu_{\mathrm{coll}}$. On the passive event,
$\mu_{\mathrm{coll}}=(1.8+o(1))m^2/N=(0.45+o(1))h\to\infty$. Chebyshev gives
$|S_{\mathrm{coll}}|\ge0.35h$ with probability $1-o(1)$. Combining this with $H_2(D)\ge0.75h$
yields $H_2(D)+b+|S_{\mathrm{coll}}|\ge m+0.10h$ certified distinct starts. Outputting any $m$ of
them succeeds.
\end{proof}

\begin{remark}[Open boundary: thin mixed-profile regime]
\label{rem:thin-boundary}
\Cref{prop:k2-counterexample} shows that $Q(m,b)$ need not be the sharp threshold when
$m-b=o(m)$ and $A_N\asymp B_N$. Determining the threshold across this boundary remains open.
\end{remark}

Finally, a uniform query batch alone can recover iterates when $b$ is much larger than the output
list.

\begin{lemma}[Query-only $k$-walks under a uniform batch]
\label{lem:query-only}
Fix $k\ge2$. Let $U$ be a uniformly random $b$-subset independent of $\pi$, and let $T=H_k(U)$.
Set $\mu_k(N,b)=N(b)_k/(N)_k$ and $\lambda_k(b)=(N-k+1)(b)_k/(N)_k$. Then for $N\ge2k$ and
$2k\le b\le N$,
\[
\E[T]=\lambda_k(b)+\sum_{\ell=1}^{k-1}\frac{(b)_\ell}{(N)_\ell},\qquad
\Var(Z_k(U))\le(4k-2)\lambda_k(b)+B_k\frac{\lambda_k(b)^2}{N},
\]
and $\Prob[T\ne Z_k(U)]\le(k-1)b/N$. If $b=o(N)$ and $\mu_k(N,b)\to\infty$, then
$T=(1+o(1))\mu_k(N,b)$ with probability $1-o(1)$. In particular, a query-only learner succeeds on
$\CompRel_k(N,m_N,b)$ with probability $1-o(1)$ whenever $b=o(N)$ and $b^k/N^{k-1}\gg m_N$.
\end{lemma}

\begin{proof}
Apply \Cref{lem:prefix-moments} to $Z_k(U)$ and \Cref{lem:short-cycles} to $T-Z_k(U)$. Chebyshev
concentration then gives $Z_k(U)=(1+o(1))\lambda_k(b)$, while $\lambda_k(b)=(1-o(1))\mu_k(N,b)$
and $T=Z_k(U)$ with probability $1-o(1)$. The final condition implies
$\mu_k(N,b)=(1-o(1))b^k/N^{k-1}\gg m_N$, so selecting any $m_N$ revealed starts succeeds.
\end{proof}

\section{Related Work and Model Boundary}
\label{sec:related}

Our problem combines random passive edges, one nonadaptive query batch, and a growing list of
composed iterate certificates. Each neighboring literature differs in at least two of access
interface, hidden object, and output objective (\Cref{tab:model-comparison}).

\paragraph{Interactive learning and membership queries.}
Query-assisted learning separates passive examples from synthesized membership queries across
concept classes \citep{gur2024interactive,lange2026membership,wang2026active,diakonikolas2024halfspaces,tadepalli1998structured}.
These works evaluate hypothesis risk or testing error. Our hidden object is a single uniform
permutation, and success is exact recovery of a list of composed relations rather than learning a
predictor.

\paragraph{Pointer chasing and round complexity.}
Pointer chasing measures communication or rounds for composing functions split among parties
\citep{nisan1991rounds,ponzio2001pointer,yehudayoff2020pointer,mao2025gadgetless,fischer2025pointer}.
Our learner sees random edges of one permutation and certifies a growing list of starts. The binding
constraint is passive prefix supply and batch collisions, not communication bandwidth.

\paragraph{Hidden permutations and permutation sampling.}
Full permutation identification has been studied under score queries \citep{afshani2013hidden}, while
compressed permutation oracles address quantum predicate search and cycle finding
\citep{carolan2026compressed,carolan2026revisited}. Other work samples permutations from local
randomness or cell probes \citep{viola2020sampling,alekseev2026sampling,huang2026parallel}. Our
learner observes partial edges of an already drawn permutation and recovers only a sublinear list of
iterate pairs, not the full permutation or a fresh sample.

\paragraph{Compositional learning and batch adaptivity.}
\citet{wang2025compositional} study compositions of input and hidden permutations through
statistical-query and transformer-learning analyses. \citet{haghtalab2025adaptivity} characterize
sample-adaptivity tradeoffs for on-demand sampling across distributions. Neither model supplies the
membership-query list objective or the passive-prefix/collision accounting studied here.

\begin{table}[H]
\centering
\small
\setlength{\tabcolsep}{4pt}
\caption{Structural comparison with neighboring models. Each adjacent line differs from
$\CompRel_k(N,m,b)$ in at least two of access interface, hidden object, and output objective.}
\label{tab:model-comparison}
\begin{tabular}{@{}
  >{\raggedright\arraybackslash}p{0.25\textwidth}
  >{\raggedright\arraybackslash}p{0.23\textwidth}
  >{\raggedright\arraybackslash}p{0.22\textwidth}
  >{\raggedright\arraybackslash}p{0.23\textwidth}@{}}
\toprule
\textbf{Line of work} & \textbf{Access interface} & \textbf{Hidden object} & \textbf{Output objective}\\
\midrule
MQ / active learning \citep{gur2024interactive,diakonikolas2024halfspaces} &
examples and membership queries & concept class & low-risk hypothesis or test\\
Pointer chasing \citep{nisan1991rounds,mao2025gadgetless} &
multi-party rounds & split functions & one composed pointer\\
Permutation identification \citep{afshani2013hidden,carolan2026compressed} &
score, quantum, or cell-probe & full $\pi\in\Perm_N$ & full $\pi$ or a fresh sample\\
Compositional learning \citep{wang2025compositional} &
statistical queries & $k$ hidden permutations & composed predictor\\
\textbf{This paper} & i.i.d.\ edges and one batch & uniform $\pi\in\Perm_N$ &
$m$ distinct-start $\pi^k$ pairs\\
\bottomrule
\end{tabular}
\end{table}

\section{Discussion and Open Problems}
\label{sec:discussion}

The sharp threshold is governed by a two-population supply bottleneck, not by the absence of batch
collisions. Retaining $\Ccoll$ in the deterministic walk-capacity inequality separates the failure
of a zero-collision proof from an actual change in sample complexity.

\begin{remark}[Zero-collision union bound diagnostic]
\label{rem:zero-collision}
Fix $k\ge2$ and positive integer sequences $h_N\le m_N=o(N)$, and set $b_N=m_N-h_N$. If
$Q_{k-1}(N)\ge Q_k(N)$, then $m_N^k\ge h_N^{k-1}N$, hence
$(h_N/m_N)^{k-1}\le m_N/N=o(1)$. Since $k$ is fixed, $h_N=o(m_N)$ and
\[
\frac{b_NQ_{k-1}(N)}N
=(1-o(1))\left(\frac{m_N^k}N\right)^{1/(k-1)}
\ge(1-o(1))h_N.
\]
Thus the usual union bound for a collision-free batch fails on the short-prefix branch. This is a
failure of that proof technique, not a second phase transition; an arbitrary learner may always
submit an empty batch.
\end{remark}

\paragraph{Limitations and model boundaries.}
The results are distributional, information-theoretic, and restricted to fixed depth $k$, sublinear
lists $m=o(N)$, and one nonadaptive batch. Four concrete extensions remain open.

\begin{enumerate}[label=\textbf{\arabic*.}]
\item \emph{The thin mixed-profile boundary.} When $m-b=o(m)$ and $A_N\asymp B_N$,
\Cref{prop:k2-counterexample} shows that nonterminal collisions can beat the naive two-scale value.
A complete threshold requires optimizing every decomposition of a walk into passive segments and
colliding query edges.
\item \emph{Multi-round query adaptivity.} A second batch can query endpoints revealed by the first,
so the capacity inequality no longer applies directly. The sharp sample-round tradeoff for
$r\ge2$ batches remains open.
\item \emph{Critical window, linear lists, and growing depth.} The present statements do not resolve
$q=(1+o(1))Q(N)$, $m=\Theta(N)$, or $k=k(N)\to\infty$. These regimes require concentration bounds
with explicit dependence on $k$ and on the distance to the threshold.
\item \emph{Sampling and computation.} A learner-chosen list of valid pairs does not sample the full
graph law of $\pi^k$ in total variation, and no polynomial-time or statistical-query restriction is
imposed. Those objectives require separate lower-bound analyses.
\end{enumerate}

\label{mainbody:end}
\typeout{MAIN_BODY_LAST_PAGE=\thepage}

\clearpage
\fancyhead[R]{\PaperRightHeader{References}}
\markboth{}{References}
\bibliographystyle{plainnat}
\bibliography{comprel_refs}

\clearpage
\appendix
\fancyhead[R]{\PaperRightHeader{\rightmark}}
\markboth{}{Appendix}
\providecommand{\theHsection}{}\renewcommand{\theHsection}{appendix.\Alph{section}}
\providecommand{\theHsubsection}{}\renewcommand{\theHsubsection}{\theHsection.\arabic{subsection}}
\providecommand{\theHtheorem}{}\renewcommand{\theHtheorem}{\theHsection.\arabic{theorem}}
\providecommand{\theHequation}{}\renewcommand{\theHequation}{\theHsection.\arabic{equation}}
\providecommand{\theHtable}{}\renewcommand{\theHtable}{\theHsection.\arabic{table}}
\providecommand{\theHfigure}{}\renewcommand{\theHfigure}{\theHsection.\arabic{figure}}
\section{Algebraic Checks, Packing Procedure, and Finite Witness}
\label{app:checks}

This appendix supplies the combinatorial lemmas used by the main-body moment bounds, the explicit
packing learner, and an exact finite-instance consistency check. \Cref{tab:appendix-map} records the
dependency map. All main theorems and propositions are proved in \Cref{sec:proofs,sec:decoupled}.

\begin{table}[htb]
\centering
\small
\caption{Dependency map for the appendix.}
\label{tab:appendix-map}
\begin{tabular}{@{}lll@{}}
\toprule
\textbf{Subsection} & \textbf{Role} & \textbf{Main-body statements supported}\\
\midrule
\Cref{app:algebra} & Combinatorial ingredients & \Cref{lem:prefix-moments,lem:short-cycles,lem:query-only}\\
\Cref{app:algorithm} & Explicit packing procedure & \Cref{thm:two-branch,cor:passive-baseline,thm:decoupled} (upper bounds)\\
\Cref{app:finite-diagnostic} & Exact finite witness & Consistency check of \Cref{def:task} at $k=3$, $m=2$\\
\bottomrule
\end{tabular}
\end{table}

\subsection{Elementary Combinatorial Lemmas for Prefix Moments}
\label{app:algebra}

\begin{lemma}[Uniform cycle-length distribution]
\label{lem:cycle-length}
Fix $N\ge1$, $r\in[N]$, and $\ell\in[N]$. If $\pi\sim\Unif(\Perm_N)$ and $L(r)$ is the cycle length
containing $r$, then $\Prob[L(r)=\ell]=1/N$.
\end{lemma}

\begin{proof}
The ordered orbit tuple $(\pi(r),\dots,\pi^{\ell-1}(r))$ consists of $\ell-1$ distinct points other
than $r$, followed by a return to $r$. There are $(N-1)_{\ell-1}$ choices for that tuple and
$(N-\ell)!$ completions on the remaining points. Their product is $(N-1)!$, so division by $N!$
gives $1/N$.
\end{proof}

\begin{lemma}[Falling-factorial variance ratio]
\label{lem:falling-factorial-ratio}
For every integer $a\ge1$, there is a finite $B_a>0$ such that for all integers $N\ge2a$,
\begin{equation}\label{eq:falling-ratio}
R_a(N)\defeq\frac{N^2}{(N-a+1)^2}\frac{(N)_a}{(N-a)_a}\le1+\frac{B_a}{N}.
\end{equation}
\end{lemma}

\begin{proof}
For $N\ge2a$, every denominator in \eqref{eq:falling-ratio} is positive. If $N\ge4a^2$, then
$N-a-i\ge N/2$ for every $0\le i<a$, so
\[
\frac{(N)_a}{(N-a)_a}
=\prod_{i=0}^{a-1}\left(1+\frac a{N-a-i}\right)
\le\left(1+\frac{2a}N\right)^a.
\]
Set $u=2a^2/N\le1/2$. Binomial expansion and $\binom a j\le a^j$ give
\[
\left(1+\frac{2a}N\right)^a
\le1+\sum_{j=1}^a\left(\frac{2a^2}N\right)^j
\le1+\frac u{1-u}\le1+\frac{4a^2}N.
\]
Also $N/(N-a+1)\le1+2a/N$, hence its square is at most $1+8a/N$. Therefore
\[
R_a(N)\le(1+8a/N)(1+4a^2/N)
\le1+\frac{32a^3+4a^2+8a}{N}.
\]
For the finitely many integers $2a\le N<4a^2$, $R_a(N)$ is finite. Taking $B_a$ to be the maximum
of $32a^3+4a^2+8a$ and $\max_{2a\le N<4a^2}N(R_a(N)-1)$ completes the proof.
\end{proof}

\subsection{Explicit Mixed-Prefix Packing Procedure}
\label{app:algorithm}

\Cref{alg:mixed-packing} records the deterministic learner used in the upper bounds.

\begin{algorithm}[H]
\caption{Mixed-Prefix Packing Learner for $\CompRel_k(N,m,b)$}
\label{alg:mixed-packing}
\begin{paperalgorithmic}
\Require Depth $k\ge2$, domain size $N\ge m\ge1$, batch budget $b\in\Zge$, truncation $q_0\in\Zge$,
and passive stream $((X_i,\pi(X_i)))_{i=1}^q$ with $q\ge q_0$.
\Ensure A list $L=((R_1,C_1),\dots,(R_m,C_m))$.
\State Set $\bact\gets\min(m,b)$ and $h\gets(m-b)_+$.
\State Truncate to the first $q_0$ samples; set $D\gets\{X_1,\dots,X_{q_0}\}$ and $P\gets\pi|_D$.
\State Compute $\Zset_k(D)$ and $\Zset_{k-1}(D)$ from $P$.
\If{$|\Zset_k(D)|\ge h$ and $|\Zset_{k-1}(D)\setminus\Zset_k(D)|\ge\bact$}
\State Choose $h$ starts $S_{\mathrm{free}}\subseteq\Zset_k(D)$ and $\bact$ starts
$S_{\mathrm{paid}}\subseteq\Zset_{k-1}(D)\setminus\Zset_k(D)$.
\State Submit the nonadaptive batch
$\Uall\gets\{P^{k-1}(r):r\in S_{\mathrm{paid}},\,P^{k-1}(r)\notin D\}$ and observe $\pi|_{\Uall}$.
\State Set $M\gets P\cup\pi|_{\Uall}$.
\For{each $r\in S_{\mathrm{free}}$} \State Append $(r,P^k(r))$ to $L$. \EndFor
\For{each $r\in S_{\mathrm{paid}}$} \State Append $(r,M(P^{k-1}(r)))$ to $L$. \EndFor
\State \Return $L$.
\Else
\State Submit the empty batch and \Return $((1,1),(2,2),\dots,(m,m))$.
\EndIf
\end{paperalgorithmic}
\end{algorithm}

\begin{proposition}[Deterministic invariants of \Cref{alg:mixed-packing}]
\label{prop:packing-deterministic}
The algorithm submits at most $b$ queries and outputs $m$ pairs with distinct starts. Whenever
$Z_k(D)\ge h$ and $|\Zset_{k-1}(D)\setminus\Zset_k(D)|\ge\bact$, it succeeds on
$\CompRel_k(N,m,b)$.
\end{proposition}

\begin{proof}
The two selected start sets are disjoint and have total size $h+\bact=m$. For each paid start,
$P^{k-1}(r)=\pi^{k-1}(r)$ is known and distinct by injectivity, so at most $\bact\le b$ queries are
used. Free starts satisfy $P^k(r)=\pi^k(r)$, and paid starts satisfy
$M(P^{k-1}(r))=\pi^k(r)$. The fallback list also has $m$ distinct starts and uses no queries.
\end{proof}

\subsection{Exact Finite Witness (Non-Evidentiary Bookkeeping Diagnostic)}
\label{app:finite-diagnostic}

The companion files \texttt{diagnostics/exact\_list\_dp.py} and \texttt{diagnostics/results.csv}
enumerate all permutations and passive input sequences for $\CompRel_3(N,2,b)$ with
$N\in\{4,5,6\}$, $q\in\{0,1,2,3\}$, and $b\in\{0,1\}$. The entries below are exact rational
Bayes-optimal success probabilities. One passive edge and one blind query induce the same posterior,
which explains the repeated entries. These values check the finite objective only; they do not
estimate the asymptotic threshold.

\begin{table}[htb]
\centering
\footnotesize
\setlength{\tabcolsep}{5pt}
\caption{Exact optimal success probabilities for $\CompRel_3(N,2,b)$. \emph{Evidence status:} exact
finite witness, non-evidentiary bookkeeping diagnostic.}
\label{tab:finite-diagnostic}
\begin{tabular}{@{}ccrrr@{}}
\toprule
$N$ & $q$ & \multicolumn{1}{c}{$b=0$} & \multicolumn{1}{c}{$b=1$} & \multicolumn{1}{c}{$\Delta$}\\
\midrule
4 & 0 & $5/12$ (0.4167) & $13/24$ (0.5417) & $+0.1250$\\
4 & 1 & $13/24$ (0.5417) & $7/8$ (0.8750) & $+0.3333$\\
4 & 2 & $73/96$ (0.7604) & $31/32$ (0.9688) & $+0.2083$\\
4 & 3 & $337/384$ (0.8776) & $127/128$ (0.9922) & $+0.1146$\\
\addlinespace
5 & 0 & $1/4$ (0.2500) & $3/10$ (0.3000) & $+0.0500$\\
5 & 1 & $3/10$ (0.3000) & $11/20$ (0.5500) & $+0.2500$\\
5 & 2 & $1/2$ (0.5000) & $71/100$ (0.7100) & $+0.2100$\\
5 & 3 & $153/250$ (0.6120) & $431/500$ (0.8620) & $+0.2500$\\
\addlinespace
6 & 0 & $1/5$ (0.2000) & $89/360$ (0.2472) & $+0.0472$\\
6 & 1 & $89/360$ (0.2472) & $13/30$ (0.4333) & $+0.1861$\\
6 & 2 & $809/2160$ (0.3745) & $299/540$ (0.5537) & $+0.1792$\\
6 & 3 & $6089/12960$ (0.4698) & $409/540$ (0.7574) & $+0.2876$\\
\bottomrule
\end{tabular}
\end{table}

\end{document}